\documentclass{physeauth}
\usepackage{graphicx}
\usepackage{amsmath}
\usepackage{amssymb}
\usepackage{dsfont}

\newcommand{\eq}[1]{Eq.~(\ref{#1})}

\newcommand{\lapx}{\,\raisebox{-.5ex}
 {$\stackrel{\raisebox{-.5pt}{$\textstyle <$}}{\sim}$}\,}

\newcommand{\rme}{{\rm e}}
\newcommand{\rmi}{{\rm i}}

\begin{document} 

\begin{frontmatter}

\title{Dynamics of coupled spins in the white- and quantum-noise regime}

\author[stu]{Peter N\"agele} and
\author[stu]{Ulrich Weiss \thanksref{thank1}}

\address[stu]{II. Institut f\"ur Theoretische Physik, Universit\"at Stuttgart,
 D-70550 Stuttgart, Germany.}

\thanks[thank1]{
Corresponding author. 
E-mail: weiss@theo2.physik.uni-stuttgart.de}

\begin{abstract}
We study the dynamics of dissipative spins for general spin-spin coupling. 
We investigate the population dynamics and relaxation of the purity in the white noise regime, in which exact 
results are available. Inter alia, we find distinct reduction of decoherence and slowdown of purity decay
around degeneracy points. We also determine in analytic form the one-phonon exchange contribution to decoherence
and relaxation in the ohmic quantum noise regime valid down to zero temperature.
\end{abstract}

\end{frontmatter}


\section{Introduction}

The spin-boson model is a key model since the 80's for the quantitative study of decoherence, relaxation and energy dissipation 
\cite{leggett,bookweiss}. In view of the substantial progress in fabrication of coupled qubit devices 
\cite{pashkin_quantum_2003,clarke}
and major advances in quantum state manipulation towards quantum computation \cite{nielsen-chuang}, 
there is growing interest in the accurate calculation of the dynamics of coupled spins for realistic environmental couplings.

Here we communicate new analytical results for the dynamics of two mutually interacting spins each liable to independent white noise
or to quantum noise forces. Previous work \cite{naegele} is extended in three different directions. First, we determine the dynamics of the purity
in the white noise regime (WNR). Second, we study the dynamics near two different degeneracy points and find striking reduction of
purity decay and decoherence. Third, we calculate the exact one-phonon contribution to dephasing and relaxation in the quantum noise
regime in analytic form. The resulting expressions hold down to zero temperature. 
The major results are obtained within the real-time path sum method for the 16 states of the two-spin density matrix.
The environmental couplings are included via the Feynman-Vernon method. Here we omit methodical and technical aspects. We rather
put emphasis on results in analytic form and their physical implications

In section \ref{smodel} the model of two interacting spins each coupled to its own dissipative environment is introduced. 
An analytic analysis of the white noise regime, inter alia study of purity and degeneracy points, is given in section \ref{swnr},
while section \ref{sqnr} deals with the effects of weak quantum noise. 

\section{Model} \label{smodel}
We consider two two-state systems or spins which are mutually coupled via Ising, XY, and/or Heisenberg coupling.
In addition, each of them is coupled to its own heat bath. In pseudospin representation, the two-spin-boson Hamiltonian reads 
(we put $\hbar =k_{\rm B}^{} =1$)
\begin{equation} \label{ham1}
H = H_\mathrm{SS} + H_\mathrm{SR}  \; .
\end{equation}
Here, $H_\mathrm{SS}$ represents the interacting two spins,
\begin{equation}\label{ham2}
\begin{array}{rcl}
H_\mathrm{SS} &=& - {\textstyle \frac{1}{2}}\Delta_1\, \sigma_x - {\textstyle \frac{1}{2}} \Delta_2 \, \tau_x 
- {\textstyle \frac{1}{2}} \epsilon_1 \,\sigma_z - {\textstyle \frac{1}{2}} \epsilon_2\, \tau_z \, \\
 && -\; {\textstyle \frac{1}{2}} v_x \,\sigma_x \tau_x - {\textstyle \frac{1}{2}} v_y \,\sigma_y \tau_y 
-  {\textstyle \frac{1}{2}} v_z\,\sigma_z \tau_z \; .
\end{array}
\end{equation}
In the basis formed by the localized  states
$|R\!\!>$ and $|L\!\! >$, the parameters $\epsilon_{1,2}^{}$ and $\Delta_{1,2}$ represent the bias energies and tunneling couplings of the $\sigma$- and $\tau$-spin, and $v_{x,y,z}$ are the interaction parameters.
The term $H_\mathrm{SR}$ describes the spin-reservoir couplings and the reservoirs,
\[
H_\mathrm{SR} = -\,\frac{1}{2}\sigma_z X_1 \,-\,\frac{1}{2}\tau_z X_2 
+  \sum_{\zeta=1,2}\sum_\alpha \omega_{\zeta,\alpha}^{} b_{\zeta,\alpha}^\dag b_{\zeta,\alpha}^{} \; .
\]
Here, 
$X_\zeta(t) = \sum_{\alpha} c_{\zeta,\alpha}^{} [\,b_{\zeta,\alpha}^{}(t)\,+\,b_{\zeta,\alpha}^\dag(t)\,] \;(\zeta = 1,2)$
is a collective reservoir mode. All effects of the environment are carried by the power spectrum of the 
collective bath modes. We have $S_{\zeta,\zeta'}^{}(\omega) =  \delta_{\zeta,\zeta'}^{} S_\zeta^{}(\omega)$, where
\begin{equation}\label{powerspec}
\begin{split}
S_{\zeta}^{}(\omega) \,&=\, {\rm Re}\! \int_{-\infty}^\infty\!\!\!{\rm d}t\, \rme^{\rmi \omega t}_{} \!
\left\langle X_{\zeta}^{}(t)X_{\zeta}^{}(0)\,\right\rangle_{\!\beta} \\
& =\, \pi G_\zeta^{}(\omega)\coth\Big(\!\frac{\omega}{2 T}\!\Big) \; .
\end{split}
\end{equation}
The spectral density of the coupling is \cite{leggett,bookweiss}
\[
G_\zeta(\omega)\;=\; \sum_\alpha  c^2_{\zeta\!,\,\alpha} \delta(\omega - \omega_{\zeta\!,\,\alpha} )
\;=\; 2 K_\zeta\, \omega \,{\rm e}^{-|\omega| /\omega_{\rm c}^{} }_{} \; .
\]
The second form describes ohmic coupling with high-frequency cut-off $\omega_{\rm c}^{}$ and dimensionless
damping constant $K_\zeta$. Instead of the independent baths, one might also choose a common bath for the 
two spins \cite{wilhelm}.

In the sequel, we confine ourselves to the case of $v_y^{}$- and $v_z^{}$-coupling of the two spins.
This case is most interesting concerning application to coupled Josephson junctions. 
In addition, we disregard the bias terms. 

The Hamiltonian $H_\mathrm{SS}$ is diagonalized with the unitary matrix
\[
U=\frac{1}{2}\left(
\begin{array}{llll}
 \frac{\cos \left(\phi _1\right)}{f_+(\phi_1)} & f_+(\phi_1) & f_+(\phi_1) & \frac{\cos \left(\phi _1\right)}{f_+(\phi_1)} \\[2mm]
 -\frac{\cos \left(\phi _2\right)}{f_-(\phi_2)} & f_-(\phi_2) & - f_-(\phi_2) & \frac{\cos \left(\phi _2\right)}{f_-(\phi_2)} 
\\[2mm]
 -\frac{\cos \left(\phi _2\right)}{f_+(\phi_2)} & -f_+(\phi_2) & f_+(\phi_2) & \frac{\cos \left(\phi _2\right)}{f_+(\phi_2)} 
\\[2mm]
 \frac{\cos \left(\phi _1\right)}{f_-(\phi_1)} & -f_-(\phi_1) & -f_-(\phi_1) & \frac{\cos \left(\phi _1\right)}{f_-(\phi_1)}
\end{array}
\right) ,
\]
with $f_{\pm}(\phi) = \sqrt{1 \pm \sin \phi}\; $ and mixing angles
$\phi_{1,2} = \arctan\left[ (v_y \mp v_z)/(\Delta_1\pm\Delta_2) \right]$. We then get
\begin{equation}
\widetilde H = U\,H\,U^{-1} = -\frac{\Omega}{2} \,(\sigma_z \otimes \mathds{1}) - \frac{\delta}{2} \, (\mathds{1} \otimes \tau_z) ,
\end{equation}
with the eigen frequencies
\begin{equation}\label{barefreq}
 \begin{split}
 \Omega &= {\textstyle \frac{1}{2}}( \Omega_{+}  + \Omega_{-} )\, , \qquad \, \delta = {\textstyle \frac{1}{2}}
( \Omega_{+} - \Omega_{-} ) \, , \\[1mm]
 \Omega_{\pm} &= \sqrt{(\Delta_1 \pm \Delta_2)^2 + (v_y \mp v_z)^2}\; .
 \end{split}
\end{equation}
We have the Vieta relations
\begin{equation}
 \begin{split}
  \Omega_{+}^{2} + \Omega_{-}^{2} &= 2\,( \Delta_1^2 + \Delta_2^2 + v_y^2 + v_z^2 ) \; , \\
  \Omega^2+\delta^2 &= \Delta_1^2 + \Delta_2^2 + v_y^2  + v_z^2 \; ,\\
  \Omega^2 \, \delta^2 &= (v_y\,v_z - \Delta_1\,\Delta_2)^2\; .
 \end{split}
\end{equation}

The two-spin density matrix has 16 matrix elements. They can be expressed as linear combinations of the unit matrix 
and the following 15 expectation values, $\langle \sigma_i\otimes \mathbf{1} \rangle_t \equiv \langle \sigma_i\rangle_t$, 
$\langle \mathbf{1} \otimes \tau_i \rangle_t \equiv \langle \tau_i\rangle_t $, and  
$\langle \sigma_i \otimes \tau_j \rangle_t \equiv \langle \sigma_i \tau_j \rangle_t$ ($i=x,\,y,\,z$ and $j=x,\,y,\,z$). These 
quantities, denoted by $W_j(t)$ ($j=1,\cdots,\,15$), obey the  equations of motion
$\dot W_j(t) = \,\mathrm{i}\, [\,H_\mathrm{SS},\,W_j(t)\,]$ ($j=1,\cdots,15$). The set of coupled equations are conveniently solved in Laplace space $\lambda$. Throughout we choose the initial state $\langle \sigma_z\rangle_0 =1$, $\langle \tau_z\rangle_0^{}=1$,
and $\langle \sigma_z \tau_z \rangle_0^{}=1$, and all other expectations zero.

The resulting expressions may be written as
\begin{equation}\label{fracexpr}
W_j(\lambda) = N_j^{}(\lambda)/D_j^{}(\lambda) \; ,\quad j=1,\cdots,\,15 \; ,
\end{equation}
where the $N_j^{}(\lambda)$  are different for the individual $W_j(\lambda)$, while the 
$D_j^{}(\lambda)$ fall into the following two categories, 
\begin{equation}
\begin{array}{rcl}
D_{\Omega,\delta}(\lambda) &=& (\lambda^2 +\Omega^2)(\lambda^2+\delta^2) \;,\\
D_{\Omega_\pm}(\lambda) &=& \lambda(\lambda^2 +\Omega^2_+)(\lambda^2+\Omega^2_-) \; .
\end{array}
\end{equation}
\section{White-noise regime}\label{swnr}
\subsection{General features and qualitative behavior}\label{ssgf}
The exact formal solution of the dissipative two-spin dynamics has been discussed in Ref. \cite{naegele}. Explicit
expressions for the $W_j(\lambda)$ have been given in the white noise regime (WNR), in which (\ref{powerspec}) reduces to
\begin{equation}\label{wnl}
S_\zeta(\omega \ll T ) = 2\vartheta_\zeta\;,  \quad\mbox{where}\quad \vartheta_\zeta = 2 \pi K_\zeta T  \; .
\end{equation}
It was found that the form (\ref{wnl}) is expedient in the regime $K_\zeta\lapx 0.3$ and $\Omega_\pm\lapx T\ll\omega_{\rm c}$, which covers
not only the incoherent regime but also a sizeable domain of the coherent regime. The analysis shows that reservoir modes in the range
$2\pi T < \omega <\omega_{\rm c}^{}$ give rise to an adiabatic (Franck-Condon-type) renormalization of the tunneling coupling,
$\Delta_\zeta^2 \to\bar\Delta_\zeta^2 = (2\pi T/\Delta_{\zeta,{\rm r}})^{2K_\zeta}_{}\Delta_{\zeta,{\rm r}}^2$ with
$\Delta_{\zeta,{\rm r}}^{1-K_\zeta}=\Delta_\zeta/\omega_{\rm c}^{K_\zeta}$
\cite{bookweiss,naegele}. 
In the reminder of this section, we assume that the $\Delta_{1,2}$ are the renormalized ones.
The modes with $\omega<2\pi T$ lead to decoherence and relaxation. In the WNR, they are accounted for
by an appropriate shift of the Laplace variable $\lambda$ in the time interval, in which spin $\zeta$ dwells in an 
off-diagonal state. We have ($\zeta=1,2$)
\[
\lambda\to\lambda_\zeta = \lambda + \vartheta_\zeta\;,\quad\mbox{and}\quad 
\lambda \to\lambda_{12} = \lambda + \vartheta_1 + \vartheta_2 \; .
\]
The resulting 15  coupled equations are
\[
\begin{array}{rcl}
\lambda\,  \langle\sigma _z\rangle &=& 1 - \Delta _1 \langle\sigma _y\rangle - v_x \langle\tau _x \sigma _y\rangle + v_y \langle\sigma _x \tau _y\rangle \, , \\
\lambda_1\, \langle\sigma _y\rangle &=& \Delta _1 \langle\sigma _z\rangle 
+ v_x \langle\sigma _z \tau _x\rangle - v_z \langle\tau _z \sigma_x\rangle  \, , \\ 
\lambda_1\, \langle\sigma _x\rangle &=& 
- v_y \langle\sigma _z \tau _y\rangle + v_z \langle\sigma _y \tau _z\rangle \, , \\ 
\lambda\,  \langle\tau_z\rangle &=& 1 - \Delta _2 \langle\tau _y\rangle - v_x \langle\sigma _x \tau _y\rangle + v_y \langle\sigma _y \tau _x\rangle \, , \\ 
\lambda_2\, \langle\tau _y\rangle &=& \Delta _2 \langle\tau _z\rangle 
+ v_x \langle\sigma _x \tau_z\rangle - v_z \langle\sigma _z \tau _x\rangle \, , \\ 
\lambda_2\, \langle\tau _x\rangle &=& 
- v_y \langle\sigma _y \tau _z\rangle + v_z \langle\sigma _z \tau _y\rangle \, , \\ 
\lambda_{12}\, \langle\sigma _x \tau _y\rangle &=& \Delta _2 \langle\sigma _x \tau _z\rangle 
+\, v_x \langle\tau _z\rangle - v_y \langle\sigma _z\rangle \, , \\ 
\lambda_{12}\, \langle\sigma _y \tau_x\rangle &=&  \Delta _1 \langle\tau _x \sigma _z\rangle 
+\,  v_x \langle\sigma _z\rangle - v_y \langle\tau _z\rangle \, , \\ 
\lambda_2\langle\sigma _z \tau _y\rangle &=& \Delta _2 \langle\sigma _z \tau _z\rangle \!-\! \Delta _1 \langle\sigma _y \tau_y\rangle 
\!+\!  v_y \langle\sigma _x\rangle \!-\! v_z \langle\tau _x\rangle  \, , \\ 
\lambda_1 \langle\sigma _y \tau _z\rangle &=& \Delta _1 \langle\sigma _z \tau _z\rangle \!-\! \Delta _2 \langle\sigma _y \tau _y\rangle 
\!+\!  v_y \langle\tau_x\rangle \! -\! v_z \langle\sigma _x\rangle \, , \\ 
\lambda_1\, \langle\sigma _x \tau _z\rangle &=&  - \Delta _2 \langle\sigma _x \tau _y\rangle 
- v_x \langle\tau _y\rangle + v_z \langle\sigma _y\rangle  \, , \\ 
\lambda_2\, \langle\sigma _z \tau _x\rangle &=& - \Delta _1 \langle\tau_x \sigma _y\rangle 
- v_x \langle\sigma _y\rangle  + v_z \langle\tau _y\rangle  \, , \\ 
\lambda_{12}\, \langle\sigma _y \tau _y\rangle &=& \Delta _1 \langle\sigma _z \tau _y\rangle + \Delta _2 \langle\sigma_y \tau _z\rangle\, , \\ 
\lambda\,  \langle\sigma _z \tau _z\rangle &=& 1 - \Delta _1 \langle\sigma _y \tau _z\rangle - \Delta _2 \langle\sigma _z \tau _y\rangle\, , \\
\lambda_{12}\, \langle\sigma _x \tau _x\rangle &=&  0 \, . 
\end{array}
\]
The solutions for all $W_j(\lambda)$ are again in the form (\ref{fracexpr}). There arise
only three different denominators, namely
\[
\begin{split}
D_1(\lambda) &= (\lambda\lambda_1+\Delta_1^2)(\lambda_{1}\lambda_{12} +\Delta_2^2) + v_y^2(\lambda_1^2+v_z^2 )  \\
    &\quad+ \lambda \lambda_{12} v_z^2 - 2 v_y v_z \Delta_1\Delta_2   \; , \\[1mm]
D_2(\lambda) &= (\lambda\lambda_2+\Delta_2^2)(\lambda_{2}\lambda_{12} +\Delta_1^2) + v_y^2(\lambda_2^2+v_z^2 )  \\
    &\quad+ \lambda \lambda_{12} v_z^2 - 2 v_y v_z \Delta_1\Delta_2   \; , \\[1mm]
D_3(\lambda) &= \lambda\lambda_{12} [\, \lambda_1^2\lambda_2^2 
+  v_z^2(\lambda_1^2+\lambda_2^2 )   +  2 v_y^2 \lambda_1\lambda_2 \, ] \\ 
&+ (v_y^2 \Delta_2^2 + v_z^2\Delta_1^2 + \Delta_2^2 \lambda_1\lambda_2 ) (\lambda_1\lambda_{12}+ \lambda\lambda_2 ) \\
&+ (v_y^2 \Delta_1^2 + v_z^2\Delta_2^2 + \Delta_1^2 \lambda_1\lambda_2 ) (\lambda_2\lambda_{12}+ \lambda\lambda_1 ) \\
&+ (v_y^2 - v_z^2)^2 \lambda\lambda_{12} + (\Delta_1^2-\Delta_2^2)^2 \lambda_1\lambda_2   \\
&+ 2v_y v_z \Delta_1\Delta_2 (\lambda+\lambda_{12})(\lambda_1+\lambda_2 ) 
   \, .
\end{split} 
\]
For instance, we obtain
\begin{equation}\label{undamp}
\begin{split}
\langle \sigma_z (\lambda) \rangle &= \frac{\lambda_1 \left( \Delta_2^2 + \lambda_1 \lambda_{12} \right) + v_z^2 \lambda_{12}}{D_1(\lambda)}
\;, \\[1mm]
\langle \tau_z (\lambda) \rangle &= \frac{\lambda_2 \left( \Delta_1^2 + \lambda_2 \lambda_{12} \right) + v_z^2 \lambda_{12}}{D_2(\lambda)} 
\; ,\\[1mm]
\langle \sigma_y\tau_y (\lambda) \rangle &= \big[\,\Delta_1\Delta_2(v_y^2\!+\! v_z^2+\lambda_1\lambda_2)\\
&\quad \;\;+ v_y v_z(\Delta_1^2\!+\! \Delta_2^2)\, \big] \,\frac{\lambda_1+\lambda_2}{ D_3(\lambda)}  \; .
\end{split}
\end{equation}
For vanishing bath coupling, both $D_{1}(\lambda)$ and $D_{2}(\lambda)$ go to $D_{\Omega,\delta}(\lambda)$, 
and $D_3(\lambda)\to \lambda$ reduces to $D_{\Omega_\pm}(\lambda)$.

\begin{figure}[tp]
\centering
\includegraphics[width=5.2 cm, height=3.5 cm]{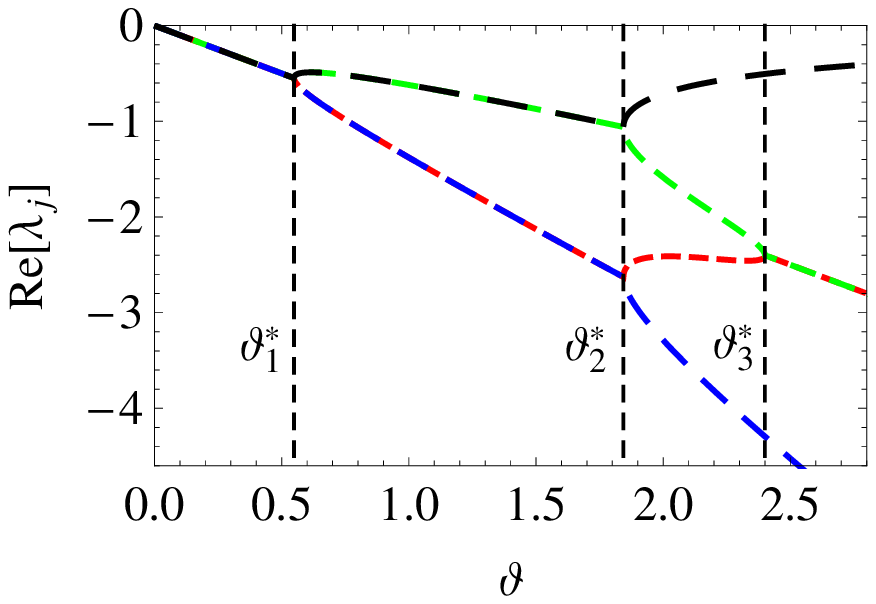}
\includegraphics[width=5.4 cm, height=3.5 cm]{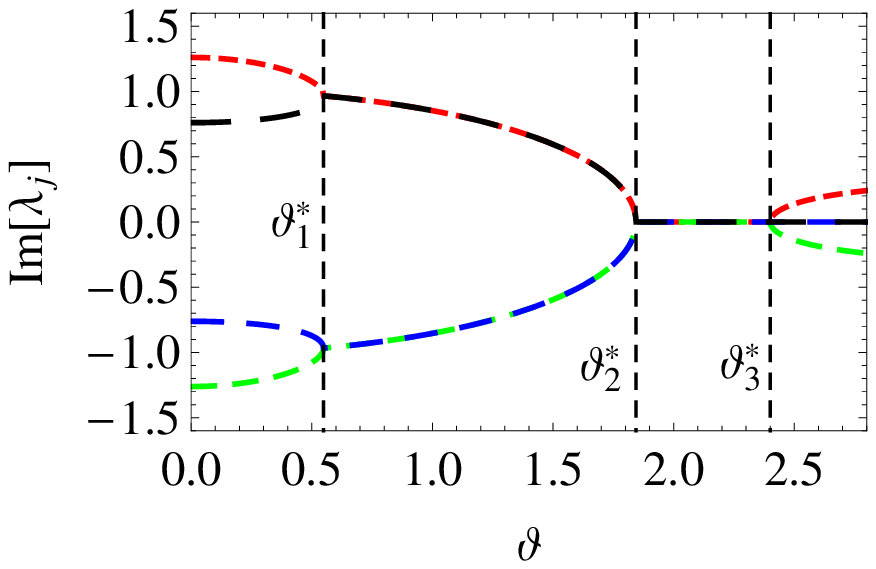}
\caption{$\langle\sigma_z\rangle$:\; Plots of ${\rm Re}\,\lambda_j$ and ${\rm Im}\,\lambda_j$ against
$\vartheta=\vartheta_{1,2}$. The parameters are $\Delta_{1,2}=1$,  $v_z=0.4$, $v_y=0.1$.  }
\label{fig:realim}
\end{figure}

\begin{figure}[tp]
\centering
\includegraphics[width=5.0 cm, height=3.5 cm]{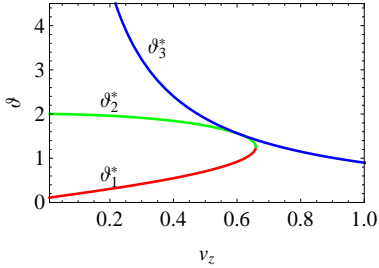}
\caption{$\langle\sigma_z^{}\rangle$: \; Crossover temperatures against $v_z$ for $v_y=0.1$ and identical spins,
$\Delta_{1,2}=1$, and $\vartheta_{1,2}=\vartheta$.}
\label{fig:sz_crosstemp}
\end{figure}

The dynamics of the expectations $A_j(t)$ is mainly determined by the zeros $\lambda_i$ of $D_k(\lambda)$ $(k=1,\,2,\,3)$.
They appear in complex conjugate or real pairs. We get
\begin{equation}
A_j(t) \;=\; \sum_{i=1}^n B_i\,{\rm e}^{\lambda_i t}_{}\;,
\end{equation}
where $n$ is either 4 or 6. 
The behaviors of the four $\lambda_j$ of $\langle\sigma_z(\lambda)\rangle$ (and of $\langle\tau_z(\lambda)\rangle$) 
and the six $\lambda_j$ of
$\langle\sigma_y\tau_y(\lambda)\rangle$, and the respective amplitudes $B_j$, are quite diversified. In Fig. \ref{fig:realim} 
we show plots of the real parts (rates) and imaginary parts (oscillation frequencies) of the four $\lambda_j$ of 
$\langle\sigma_z(\lambda)\rangle$ as functions of $\vartheta$ for a particular set of parameters (identical spins, 
$\Delta_{1,2} =\Delta,\,\vartheta_{1,2}=\vartheta$). In the coupling range $v_z < v_z^\ast$ there are
three crossover temperatures $\vartheta_1^\ast$, $\vartheta_2^\ast$ and $\vartheta_3^\ast$ at which the discriminant of $D_1(\lambda)$ 
is zero. For $v_z > v_z^\ast$ there is only a single crossover temperature $\vartheta_3^\ast$. The critical coupling strength is
\begin{equation}
v_z^\ast \;=\; {\textstyle \frac{1}{2}}\,\big[\, (2\,\Delta^2+v_y^2)^{1/2} -v_y \,\big] \; .
\end{equation} 
A plot of the crossover temperatures is shown in Fig.~\ref{fig:sz_crosstemp}.

The particular case $\Delta_1=\Delta_2$ and $v_z= -v_y$ is a degeneracy point, $\Omega=\delta$ (cf. subsection~\ref{sdegp}). 
In this case, the crossover curves $\vartheta_1^\ast(v_z)$ and $\vartheta_2^\ast(v_z)$ coincide. 

In the regime $\vartheta < \vartheta_2^\ast$ the dynamics is coherent, described by a superposition of two damped oscillations with 
amplitudes of comparable size. For
$\vartheta < \vartheta_1^\ast$, the two oscillations have different frequency, $\Omega > \delta$, but the same damping rate. In the range $\vartheta_1^\ast < \vartheta < \vartheta_2^\ast$, they have the same frequency, but different damping rates.

In the range $\vartheta > \vartheta_2^\ast$, the dynamics is incoherent with 4 different relaxation rates. The two smallest rates have sizeable
amplitudes and thus dominate the relaxation dynamics.
In the so-called Kondo regime $\vartheta >\vartheta_3^\ast$, three of the four $\lambda_j$-contributions have negligibly small 
amplitudes. The only relevant contribution is that with the smallest rate. The Kondo characteristics is that this rate
decreases with increasing temperature, $\gamma_{\rm K} =\Delta^2/\vartheta$  (cf. Fig. \ref{fig:realim}). This counter-intuitive
feature is already well-known from the ohmic single-spin model \cite{bookweiss}.


\subsection{Low temperature WN regime} \label{sys_weak_vy_vz}

The WN regime has a low temperature bound roughly given by $T \approx\Omega_{\pm}$. Above this bound and below the first crossover temperature,
$\Omega_\pm \lapx T <T_1^\ast$, the real parts of the $\{\lambda_j\}$ vary linearly with temperature. In this regime,
systematic low-temperature expansion of the zeros of $D_j(\lambda)$ ($j=1,\,2,\,3$) is straightforward. The results are as follows:\\[2mm]
$\langle \sigma_z\rangle_t$:\hspace{2mm}
There is a superposition of two damped oscillations, $\lambda_{1,2}^{} = \pm{\rm i}\,\Omega -\gamma_{\Omega}^{}$ and
$\lambda_{3,4}^{} = \pm{\rm i}\,\delta -\gamma_{\delta}^{}$. The frequencies $\Omega$ and $\delta$ are close to their bare values
given in Eq. (\ref{barefreq}) near $T=\Omega_\pm$. As $T$ is increased, they approach each other and coincide at $T=T_1^\ast$.
The respective damping rates $\gamma_\Omega^{}$ and $\gamma_\delta^{}$ in the regime $\Omega_\pm<T < T_1^\ast$ read 
\begin{equation} \label{ratesz}
\begin{split}
\gamma_\Omega^{} &= \frac{1}{2}\vartheta_1 +\frac{\Omega^2 - \Delta_1^2- v_y^2 }{2(\Omega^2_{}-\delta^2_{})}\vartheta_1 +
\frac{\Omega^2-\Delta_1^2-v_z^2}{2(\Omega^2_{}-\delta^2_{})}\vartheta_2  \, , \\[2mm]
\gamma_\delta^{} &= \frac{1}{2}\vartheta_1 + \,\frac{\Delta_1^2 + v^2_{y} - \delta^2}{2(\Omega^2_{}-\delta^2_{})}\vartheta_1 + \,
\frac{\Delta_1^2 + v_z^2  - \delta_{}^2}{2(\Omega^2_{}-\delta^2_{})} \vartheta_2 \, .
\end{split}
\end{equation}
The amplitudes of the oscillatory contributions (in zeroth order in $\vartheta_{1,2}$) read\footnote{
The amplitudes $B_{\Omega,\delta}$ and $B_{\Omega_\pm}$ are the residues of each of the corresponding two complex poles.}
\[
B_\Omega = \frac{\Omega^2 -v_z^2-\Delta_2^2}{ 2(\Omega^2-\delta^2)}\;,\quad
B_\delta = \frac{\Delta_2^2 +v_z^2-\delta^2}{ 2 (\Omega^2-\delta^2)} \;.
\]
Mutual exchanges $\Delta_{1}\leftrightarrow\Delta_{2}$ and $\vartheta_{1}\leftrightarrow\vartheta_{2}$  
yield the corresponding rates and amplitudes for $\langle\tau_z^{}\rangle_t$.\\[2mm]
$\langle \sigma_z\tau_z \rangle_t$:\hspace{2mm} 
According to the zeros of $D_3(\lambda)$, 
$\lambda_{1,2} = \pm{\rm i}\,\Omega_+  - \gamma_{\Omega_+}^{}$,  
$\lambda_{3,4} = \pm{\rm i}\,\Omega_-  - \gamma_{\Omega_-}^{}$, $\lambda_{5,6}^{} = -\gamma_{5,6}^{}$, 
there are two damped oscillatory and two relaxation contri\-bu\-tions. The frequencies $\Omega_\pm$ are close to their bare values 
given in Eq. (\ref{barefreq}) near $T=\Omega_\pm$, and they coincide at the first crossover temperature $T_1^\ast$.
The damping rates $\gamma_{\Omega_\pm}^{}$ and amplitudes $B_{\Omega_\pm}$ of the oscillations are
\begin{equation}\label{ratesztz}
 \begin{split}
\gamma_{\Omega_+}^{} &\;=\; \gamma_{\Omega_-}^{} = {\textstyle\frac{1}{2} }(\vartheta_1+\vartheta_2)\, , \\[1mm]
B_{\Omega_+} &\;=\; \frac{(\Delta_1+\Delta_2)^2}{4\Omega_+^2} \;,\quad B_{\Omega_-} = \frac{(\Delta_1 - \Delta_2)^2}{4\Omega_-^2}  \;.
 \end{split}
\end{equation}
The relaxation rates $\gamma_{5,6}^{}$ are determined by a quadratic equation, which is obtained by truncation of $D_3(\lambda)$.
The resulting expressions for the relaxation rates and associated amplitudes are
\begin{equation}\label{ratesztz1}
 \begin{split}
\gamma_{5}^{} &\;=\; \frac{\Omega^2 -\Delta_1^2 -v_y^2}{\Omega^2-\delta^2}\,\vartheta_1 
                  +  \frac{\Omega^2 -\Delta_2^2 -v_y^2}{\Omega^2-\delta^2}\,\vartheta_2 \, ,  \\[1mm]
\gamma_{6}^{} &\;=\; \frac{\Delta_1^2 +v_y^2-\delta^2}{\Omega^2-\delta^2}\,\vartheta_1 
                  \;+\;  \frac{\Delta_2^2 +v_y^2 - \delta^2}{\Omega^2-\delta^2}\,\vartheta_2 \, ,
\end{split}
\end{equation}
\[
B_{\gamma_5^{}} = \frac{(v_y\Omega + v_z\delta)^2}{(\Omega^2-\delta^2)^2}\;, \quad\;\;
B_{\gamma_6^{}} = \frac{(v_y\delta + v_z\Omega)^2}{(\Omega^2-\delta^2)^2}\;  .
\]

\subsubsection{The limit $\Delta_2\to 0$ and $v_y\to 0$}\label{sdegen}
In the limit $\Delta_2\! \to \! 0$ and $v_y\!\!\to\!\! 0$, the transition frequencies $\Omega_\pm$  become degenerate, and hence $\Omega=\sqrt{\Delta_1^2 +v_z^2}$
and $\delta =0$. 
Accordingly, the expression (\ref{ratesz}) for $\gamma_\delta^{}$ is not valid anymore. To cope with this limiting case, 
we must determine $\lambda_{3,4}$ from a quadratic equation, which is found from $D_1(\lambda) = 0 $ by reduction.
The  respective complex eigenvalues for slight detuning are found as
\[
 \lambda\; = \; \pm\, \mathrm{i} \,\sqrt{\delta^2- \frac{1}{4}\big( \frac{v_z^2}{\Omega^2}\vartheta_1   
+ \vartheta_2\big)^2\,}  - \frac{\Omega^2+\Delta_1^2}{2\Omega^2}\vartheta_1 -  \frac{\vartheta_2}{2}\; .
\]
From this form we see that the two complex conjugate eigenvalues turn into two real eigenvalues, when $\delta$ is sufficiently
small.  At $\delta=0$, the rate expressions are
\begin{equation}
\gamma_+^{} = \vartheta_1+\vartheta_2\, ,\quad \mbox{and}\quad \gamma_-^{} = \gamma_{\rm r}^{} := \frac{\Delta_1^2}{\Omega^2}\,\vartheta_1 \, .
\end{equation}
In addition, the analysis shows that the residuum associated with the pole at $\lambda=-\gamma_+^{}$ is zero, while the other yields the amplitude $B_{\gamma_-^{}}^{} = \frac{1}{2}v_z^2/\Omega^2$.

At this point, we remark that, in the limit $\Delta_2\to 0$, the coupling $v_z$ takes the role of a biasing energy 
for spin $\sigma$. Thus, the dissipative two-spin model reduces to the standard biased spin-boson model. 
The  rate $\gamma_{\rm r}^{}$ is just the relaxation rate of this model. 
Furthermore, the rate $\gamma_\Omega^{}$ in Eq. (\ref{ratesz}) reduces to the form
\begin{equation}
\gamma_\Omega^{} \;=\; {\textstyle \frac{1}{2}}\gamma_{\rm r}^{}\,+\, (v_z^2/\Omega^2)\,\vartheta_1 \; .
\end{equation} 
This expression coincides indeed with the decoherence rate of the biased spin-boson model in the WNR \cite{bookweiss}.
\subsection{Purity} \label{spur}
For a system described by the density matrix $\rho(t)$,
the purity $P(t) :=\mathrm{Tr}\rho^2(t)$ tells us whether the system is in a pure state or in a mixture.
For a pure state, there is $P=1$ while for a fully mixed state $P=1/N$. Here $N$ ist the number of the system's accessible states.
In the low temperature WN regime discusssed in the preceeding subsection, the purity $P(t)$ is found as (the index $\sigma,\tau$ refers to the respective spin)
\[
\begin{split}
&P(t) = {\textstyle \frac{1}{4} } + 
{\textstyle \frac{1}{8}}(1-C_\sigma )\,{\rm e}^{- 2\gamma_{\delta,\sigma}^{} t}_{} + 
 {\textstyle \frac{1}{8}}(1+C_\sigma )\,{\rm e}^{- 2\gamma_{\Omega,\sigma}^{} t}_{} \\
&\qquad + {\textstyle \frac{1}{8}}(1-C_\tau )\,{\rm e}^{- 2\gamma_{\delta,\tau}^{} t}_{} + 
{\textstyle \frac{1}{8}}(1+C_\tau )\,{\rm e}^{- 2\gamma_{\Omega,\tau}^{} t}_{} \\
&\qquad + {\textstyle \frac{1}{8}}(1-C_{\Omega_-}^{} )\,{\rm e}^{- 2\gamma_{\Omega_-}^{} t}_{} + 
{\textstyle \frac{1}{8}}(1-C_{\Omega_+}^{} )\,{\rm e}^{- 2\gamma_{\Omega_+}^{} t}_{} \\
&\qquad +  {\textstyle \frac{1}{4} }\frac{(v_y\Omega + v_z\delta)^2}{(\Omega^2-\delta^2)^2}\,{\rm e}^{-2\gamma_5^{}t}_{}
+  {\textstyle \frac{1}{4} }\frac{(v_y\delta + v_z\Omega)^2}{(\Omega^2-\delta^2)^2}\,{\rm e}^{-2\gamma_6^{}t}_{}
\end{split}
\]
with the amplitudes
\[
C_{\sigma,\tau}^{} = \frac{\Delta_{1,2}^2-\Delta_{2,1}^2 + v_y^2-v_z^2}{\Omega^2 -\delta^2}\, ,\quad
C_{\Omega_\pm^{}}^{} = \frac{(v_y \mp v_z)^2}{\Omega_\pm^{2}} \, .
\]
This function smoothly drops on the time-scale given by the system's damping and relaxation rates from the initial value $P(0)=1$
to the fully mixed thermal equilibrium state, $P(t\to\infty) =\frac{1}{4}$. Observe that all dephasing and relaxation rates relevant to 
the decay of the expectation values $W_j(t)$ ($j=1,\,\cdots,\,15$)  contribute to the decay of the purity.
\subsection{Decoherence dip near degeneracy points}\label{sdegp}
Of particular interest are degeneracy points of the two-spin system. There are two different cases:
\[
\begin{split}
\Delta_1=\Delta_2\, ,\quad v_y = -v_z\, \!\quad\! &\Longrightarrow \!\quad\! \;\;\;\, \Omega=\delta\; \;\;\;\,\quad\mbox{(case $I$)}\, ,\\[2mm]
v_y v_z = \Delta_1\Delta_2\, \;\;\qquad &\Longrightarrow\quad \Omega_+ =\Omega_- \, \quad\mbox{(case $II$)}\, .
\end{split}
\]
For comparison, we also study the nondegenerate point conjugate to case $I$
\[
\Delta_1=\Delta_2\, , \!\quad \!v_y = v_z\,\quad\! \Longrightarrow\quad\! \Omega,\delta=\Delta \pm v\,\quad(\mbox{case}\;\; I^\ast)  \, .
\]

Consider first $\langle\sigma_z\rangle_t$ in case $I$.
We find from Eq. (\ref{ratesz}) upon taking the limits
$\Delta_{1,2} = \lim_{\eta\to 0} \Delta\pm\frac{1}{2}\eta$ and 
$ v_{y,z} = \lim_{\kappa\to 0}\pm \;v +\frac{1}{2}\kappa$ the rate expressions
\[
\gamma_{\Omega,\delta}^{} = \frac{3}{4}\vartheta_1 + \frac{1}{4}\vartheta_2\mp \frac{\Delta}{4\Omega}(\vartheta_1+\vartheta_2)
\mp \frac{v}{4\Omega}(\vartheta_1-\vartheta_2)  \; ,
\]
where $\Omega=\sqrt{\Delta^2+v^2}$. These forms reduce for equal bath coupling $\vartheta=\vartheta_{1,2}$ to
\begin{equation}\label{sz1}
\gamma_{\Omega,\delta}^{} \;=\; \vartheta \;\mp\; \frac{\Delta}{2\Omega}\,\vartheta \; .
\end{equation}
Thus, the one rate is smaller and the other larger than $\vartheta$. 
The amplitudes associated with (\ref{sz1}) are found as
\begin{equation}\label{sz2}
B_{\Omega,\delta} \;=\;  \frac{1}{4}  \left(1 \,\pm\, \frac{\Delta +v}{\Omega } \right) \; ,
\end{equation}
Hence in the regime $v\ll\Delta$, the amplitude of the smaller rate is maximal, while that of the larger rate is negligibly small.

The results (\ref{sz1}) and (\ref{sz2}) may be compared with those of the nondegenerate case $I^\ast$. They are
\[
\gamma_{\Omega} = \gamma_\delta^{} = {\textstyle \frac{3}{4}}\,\vartheta_1 + {\textstyle \frac{1}{4}}\,\vartheta_2 \, , \quad
\mbox{and} \quad B_{\Omega}=B_\delta  ={\textstyle\frac{1}{4}}  \;  .
\]

\begin{figure}[tp]
\centering
\includegraphics[width=5.9 cm, height=3.5 cm]{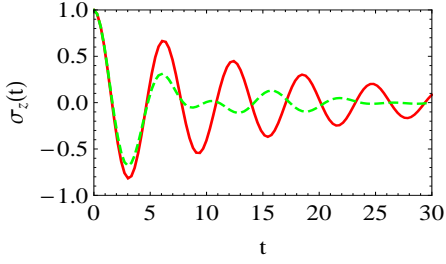}
\caption{Plots of $\langle\sigma_z^{}\rangle_t$ against $t$ for identical spins at the degeneracy point $I$ (full curve) ($v_y=-v_z$) and at the nondegenerate point $I^\ast$, $v_y=v_z$ (dashed curve). The parameters are  $v_y = 0.2$, $\Delta=1$, $K=0.01$, $T=2$.}
\label{fig:deg1}
\end{figure}

The decline of decoherence at the degeneracy point $I$ compared to point $I^\ast$ is clearly visible in Fig. \ref{fig:deg1}.
The decoherence minimum follows from competition of the two equally preferred ground states. 
Due to the $v_y$- and $v_z$-couplings, the system could relax either to parallel or to 
antiparallel alignment in $y$- or $z$-direction. Hence the 
respective second spin--spin coupling gives rise to partial suppression of  decoherence. Reduction of decoherence at point $I$
may be looked upon as a new type of frustration of decoherence \cite{novais}. Here, the phenomenon is due to
the non-commutative spin-spin couplings.

\begin{figure}[tp]
\centering
 \includegraphics[width=5.9 cm, height=3.5 cm]{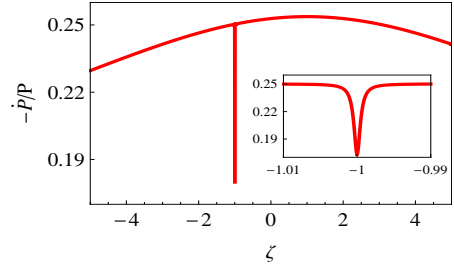}
\caption[Purity]{The quantity  $- \frac{1}{P(t)} \frac{\mathrm{d}P(t)}{\mathrm{d}t}$ is plotted against $\zeta = v_y/v_z$ for identical spins. 
There is a distinct lowering of the effective damping rate at the degeneracy point $\zeta =-1$ (case $I$).
The parameters are $\Delta=1$,\, K=0.01 and $T=3$.}
\label{fig:purimin}
\end{figure}

For $\langle \sigma_z\tau_z \rangle_t$ the picture is similar. 
In case $I$, we have $\Omega_+ = 2 \Omega$, $\Omega_- = 0$, and the rates and amplitudes read
\[
\begin{split}
\gamma_{\Omega_+}^{} &= \gamma_{\Omega_-}^{}\;=\;{\textstyle \frac{1}{2}}\,(\vartheta_1+\vartheta_2) \\
\gamma_{5,6}^{} &= 
\frac{\vartheta_1 + \vartheta_2}{2} \mp \frac{\Delta}{2\Omega}(\vartheta_1-\vartheta_2)
\mp \frac{v}{2\Omega}(\vartheta_1+\vartheta_2) \, ,  \\
B_{\Omega_+}^{} &= \frac{\Delta^2}{4\Omega^2}\, ,\quad B_{\Omega_-}^{} = \frac{1}{4}\, , \quad
B_{\gamma_5^{}} = B_{\gamma_6^{}} = \frac{v^2}{4\Omega^2} \; .
\end{split}
\]
Thus we find for equal bath couplings $\vartheta_{1,2} =\vartheta$
\begin{equation}
\gamma_{5,6}^{}\; =\; \vartheta \,\mp \frac{v}{\Omega}\,\vartheta \; .  
\end{equation}
In contrast, in the  non-degenerate case $I^\ast$ we have $\Omega_+=2\Delta$, $\Omega_- = 2 v$, 
and the rates and amplitudes are
\[
\begin{split}
\gamma_{\Omega_+}^{} &= \gamma_{\Omega_-}^{} = {\textstyle \frac{1}{2}}(\vartheta_1+\vartheta_2)\, ,\quad
\gamma_{5,6}^{} = {\textstyle \frac{1}{2}}(\vartheta_1+\vartheta_2)\, ,\\
B_{\Omega_+} &= {\textstyle \frac{1}{4}} \, , \quad B_{\Omega_-}^{} =0 \, , \quad B_{\gamma_5^{}} =B_{\gamma_6^{}} = {\textstyle
\frac{1}{4} }  \, .
\end{split}
\]
Thus we have decline of relaxation of $\langle\sigma_z\tau_z\rangle_t$ at the degeneracy point $I$.
In Fig. \ref{fig:purimin} we show a plot of $- \frac{1}{P(t)}\frac{\mathrm{d}P(t)}{\mathrm{d}t} $, 
which is a form of an effective transition rate from a pure to the fully mixed state. At the degeneracy point $\zeta =-1$ 
(case $I$), there is a distinct slowdown of the extinction  of the pure intial state. 

Case $II$ is another set of parameters for which the spectrum is degenerate, $\Omega_+ =\Omega_-$ \cite{grigorenko:040506}. 
An expedient parametrization for identical spins, $\Delta_{1,2}=\Delta$, is
\begin{equation}
\begin{split}
v_y^2 &= {\textstyle \frac{1}{2}}\, (v^2 - v_-^2)\,  , \quad v_z^2 = {\textstyle \frac{1}{2}}\, (v^2 + v_-^2)\,  ,\\
v^2 &= v_y^2 +v_z^2 \, , \quad\quad\;\; v_-^2 =\sqrt{v^4-4\Delta^4} \;>\; 0 \, .
\end{split}
\end{equation}
Then we have
\begin{equation}
  \Omega^2 \;=\; 2 \Delta^2 + v^2\;, \qquad \delta\,=\,0 \;.
\end{equation}

The undamped dynamics of $\langle\sigma_z\rangle_t$ is 
\[
\langle\sigma_z\rangle_t = \cos(\Omega t)\,+\, \frac{\Omega^2 + v_-^2}{2\Omega^2}\,\big[1\,-\,\cos(\Omega t)\big] \;,
\]
as follows from Eq.~(\ref{undamp}).
At the degeneracy point $II$, the rate expressions (\ref{ratesz}) would yield
\[
\gamma_{\Omega,\delta}^{} = \frac{1}{4}\,(3\vartheta_1+\vartheta_2) \pm \frac{v_-^2}{4\Omega^2}(\vartheta_1-\vartheta_2)\, .
\]
For identical bath couplings, these would reduce to
\begin{equation}
\gamma_\Omega^{} = \gamma_\delta^{} = \vartheta \, .
\end{equation}
Now, as we have already argued in subsection \ref{sdegen}, the expressions for $\gamma_\delta^{}$ are not correct near to 
and at the degeneracy point.
In fact, then the poles of $\langle\sigma_z(\lambda)\rangle$ at $\lambda = \pm \,{\rm i}\,\delta -\gamma_\delta^{}$ 
are determined by a quadratic equation in $\lambda$.
As one approaches the degeneracy point, the complex conjugate roots turn into two real ones. We find in the limit $\delta\to 0$
for $\vartheta_{1,2}=\vartheta$
\[
\mp \,{\rm i}\,\delta  +\gamma_\delta \;\;\to\;\; \gamma_\pm^{} = \vartheta \pm \sqrt{\frac{v^2+v_-^2}{2\Omega^2}}\vartheta
= \vartheta \pm \frac{v_z}{\Omega} \vartheta \, .
\]
Furthermore, the respective amplitudes are
\[
B_\Omega = \frac{1}{4}\,\Big( 1 - \frac{v_-^2}{\Omega^2 }\Big) \, ,\quad
B_{\gamma_\pm^{}} = \frac{1}{4}\,\Big( 1 + \frac{v_-^2}{\Omega^2} \Big) \mp \frac{1}{2} \frac{v_z}{\Omega} \, .
\]
Now, since the rate $\gamma_-^{}$ is smaller than $\vartheta$ and the respective amplitude is nonzero, we have again, 
now at point $II$, reduction of the decay of the purity $P(t)$. 

Similar behavior occurs also in $\langle\sigma_z\tau_z\rangle_t$. While $\gamma_{\Omega_+}^{}$ and $\gamma_{\Omega_-}^{}$ 
are as in Eq. (\ref{ratesztz}), we now have for $\vartheta_{1,2} =\vartheta$
\[
\gamma_5^{} = \vartheta + \frac{v_-^2}{\Omega^2}\,\vartheta\, ,\quad
\gamma_6^{} = \vartheta - \frac{v_-^2}{\Omega^2}\,\vartheta\, ,
\]
and the amplitudes read
\[
B_{\Omega_+} = \frac{\Delta^2}{\Omega^2}\,,\quad B_{\Omega_-} =0\, , \quad
B_{\gamma_{5,6}^{}} =  \frac{v^2\mp v_-^2}{2 \Omega^2}  \;. 
\]
The decrease of the rate $\gamma_{6}^{}$ for $v_-^2>0$ and of $\gamma_5^{}$ for $v_-^2>0$
leads again to a slowdown of the decay of $P(t)$ around the degeneracy point $II$. 
We have verified numerically that in the quantum noise regime $T\ll \Delta$ considered below, 
in addition to the rate $\gamma_6^{}$ or $\gamma_5$, also the dephasing rates $\gamma_{\Omega_\pm}^{}$ and $\gamma_{\Omega,\delta}^{}$ exhibit 
a pronounced minimum at the degeneracy point $II$. As a result, besides the indentation in
the purity characteristics (cf. Fig. \ref{fig:purimin}), also the dephasing  of the two-spin dynamics is considerably
slowed down at the degeneracy point $II$. A numerical analysis of this phenomenon at the degeneracy point $II$ 
is reported in Ref. \cite{grigorenko:040506}.

\section{Quantum noise regime} \label{sqnr}

Consider next the extension of the analysis to the colored quantum noise regime (QNR) relevant at $T\lapx \Omega_\pm$. 
Since at low $T$ quantum noise prevails, Eq.~(\ref{wnl})  is not valid anymore. 
Rather we have to revert to the expression (\ref{powerspec}). In the ohmic case, we have
\begin{equation} \label{powerspec1}
 S_{\zeta}(\omega) \;=\; 2\pi K_\zeta \,\omega \coth\Big(\frac{\omega}{2T}\Big) \; .
\end{equation}
We have studied the effect of the one-phonon exchange contribution to the dynamics of the two-spin model using both the perturbative
Redfield approach \cite{blum} and the self-energy method within the path sum method \cite{bookweiss}. The latter  amounts to 
systematic calculation of the self-energy to linear order in the bath correlations,
${\rm Re}\,\langle X_\zeta(t) X_\zeta(0)\rangle_T^{}= Q_\zeta(t)$  ($\zeta=1,\,2$). We have
\begin{equation}\label{bcorr}
Q_\zeta(t) \,=\, \frac{1}{\pi} \int_0^\infty \!\!{\rm d}\omega\,\,\frac{S_\zeta(\omega)}{\omega^2}\, 
[\,1-\cos(\omega t)\,] \, .
\end{equation}

In the standard notion of sojourns and blips \cite{leggett}, the one-phonon self-energy, say $\Sigma_1(\lambda)$,
receives contributions from the intra-blip correlation and from the four inter-blip correlations induced by bath 1 
between a pair of blips of the $\sigma$-spin. The inter-blip correlations vanish in the WNR. In the usual charge picture, 
the former correlation is a charge-charge interaction and the latter correlation corresponds to a dipole-dipole interaction.

In the time intervall between the correlated blips, the $\sigma$-spin may perform any number of uncorrelated jumps between
its two blip and two sojourn states. In addition, we must take into account all transitions which spin $\tau$ can make
during the dwell time of spin $\sigma$ in sojourn and blip states which are spanned by bath correlations.
The succession of flips of the $\sigma$- and $\tau$-spin is dictated by the Hamiltonian (\ref{ham1}) with (\ref{ham2}).
Following the lines expounded for the single spin-boson model \cite{bookweiss}, it is straightforward, but tedious, to calculate
the self-energy $\Sigma_1(\lambda)$. Interchange of the two spins and reservoirs then yields $\Sigma_2(\lambda)$.

The self-energies $\Sigma_{1,2}(\lambda)$ lead to shifts of the poles of the $W_j(\lambda)$. It is advantageous to
measure the resulting shifts in terms of generalized scaled temperatures $\Theta_{1,2}$. These depend on the power spectra
(\ref{powerspec}) and are normalized such that they reduce to the
previously introduced scaled temperatures $\vartheta_{1,2}$, Eq. (\ref{wnl}), in the white-noise limit. For lack of space, we now put $v_y=0$.\\[2mm]
$\langle\sigma_z\rangle_t$:\hspace{2mm}
The damping rates of the two oscillations with frequencies $\Omega$ and $\delta$ are found to read
\begin{equation} \label{sig_z_rates_low_temperature}
 \begin{split} \gamma_\Omega^{}  &= \frac{2\,\Omega^2 - \Delta_1^2- \delta^2}{2\,(\Omega^2-\delta^2)}\,\Theta_1^{(\Omega)}  
+ \frac{\Delta_2^2 -\delta^2}{2\,(\Omega^2-\delta^2)}\,\Theta_2^{(\Omega)} \, , \\
  \gamma_\delta^{} &= \frac{\Omega^2+\Delta_1^2-2\delta^2}{2\,(\Omega^2-\delta^2)}\,\Theta_1^{(\delta)} 
+ \frac{\Omega^2 -\Delta_2^2}{2\,(\Omega^2-\delta^2)}\,\Theta_2^{(\delta)}  \, ,
 \end{split}
\end{equation}
where
\[
\begin{array}{rcl}
\Theta_1^{(\Omega)} &=& \frac{\pi}{2}\, \frac{2\,(\Omega^2 - \Delta_1^2) S_1(\delta) + (\Delta_1^2 - \delta^2) S_1(\Omega)}{ 
2\Omega^2 - \delta^2 - \Delta_1^2} \; ,  \\[2mm]
\Theta_1^{(\delta)} &=& \frac{\pi}{2}\, \frac{  (\Omega^2 - \Delta_1^2) S_1(\delta) + 2(\Delta_1^2 - \delta^2) S_1(\Omega)}{\Omega^2 + \Delta_1^2 - 2\delta^2} \; .
\end{array}
\]
\begin{equation} \label{sz_gen_temp_2}
\Theta_2^{(\Omega)} = {\textstyle\frac{1}{2}}\,\pi\,S_2(\Omega) \, , \qquad\; \; \;\;
\Theta_2^{(\delta)} = {\textstyle\frac{1}{2}}\, \pi\,S_2(\delta) \, .
\end{equation}
The amplitudes asociated with the complex frequencies $\lambda=\mp\mathrm{i}\,\Omega-\gamma_\Omega^{}$ and  
$\lambda=\mp\mathrm{i}\,\delta-\gamma_\delta^{}$ are
\[
B_\Omega = \frac{\Omega^2 - \Delta_2^2 -v_z^2}{2(\Omega^2-\delta^2)} \;, \quad
B_\delta = \frac{v_z^2 +\Delta_2^2 -\delta^2}{2(\Omega^2-\delta^2)}\; .
\]
These one-phonon rate expressions hold in the QNR down to $T=0$ and they smoothly map on the WNR results (\ref{ratesz}) 
at elevated temperatures.

In the corresponding expressions for $\langle\tau_z\rangle_t$, the indices 1 and 2 are interchanged.

Following the lines expounded in subsection \ref{sdegen}, we may also consider the limit $\Delta_2\to 0$. In this limit,
the characteristics of the pole trajectories is as in subsection \ref{sdegen}. The resulting forms for $\gamma_{\rm r}^{}$
and $\gamma_\Omega^{}$ are those of the biased spin-boson model in the one-phonon QNR \cite{bookweiss},
\begin{equation}\label{sbrate}
   \gamma_{\mathrm{r}} \,= \,\frac{\pi}{2}\,\frac{\Delta_1^2}{\Omega^2} \,S_1(\Omega)\, , \quad
  \gamma = \frac{\gamma_{\mathrm{r}}}{2} \,+ \,\frac{\pi}{2}\, \frac{v_z^2}{\Omega^2}\,S_1(0)\, .
\end{equation}

Consider next the limit $v_z \rightarrow \infty$ for the symmetric case $\Delta_{1,2}=\Delta$ and $K_{1,2} =K$. For large coupling, 
the two spins are locked together and behave like a single spin with oscillation frequency $\bar \delta = \Delta^2/v_z$. 
Since the amplitude $B_\Omega$ becomes neglibly small, the dynamics is
$\langle\sigma_z^{}\rangle_t =\cos(\bar\delta t)\exp(-\gamma_{\bar\delta}^{} t)$ with the dephasing rate 
\begin{equation}\label{sbrate1}
  \gamma_{\bar\delta}^{} \;= \; \pi\, K\, \bar\delta \coth \Big( \frac{\bar\delta}{2\,T} \Big)\, ,
\end{equation}
as follows from Eq. (\ref{sig_z_rates_low_temperature}). Since the effective spin is unbiased there is no relaxation term.
In the WNR limit, the rate (\ref{sbrate1}) reduces to $\gamma_{\bar\delta}^{} = \vartheta$, which is twice the dephasing rate  
of the spin-boson model, \eq{sbrate}, in this limit. The additional factor two is because the effective spin is 
coupled to two identical reservoirs. 

The corresponding expressions for $\langle\tau_z\rangle_t$ follow from these forms by interchange of the indices 1 and 2.
\\[2mm]
$\langle\sigma_z\tau_z\rangle_t$:\hspace{2mm}
As temperature is lowered from the WNR to the QNR, the damping rates of the oscillations with frequencies $\Omega_\pm$ change from
$\gamma_{\Omega_\pm}^{} =\frac{1}{2}(\vartheta_1+\vartheta_2)$, Eq.~(\ref{ratesztz}), to the one phonon expression
\[
\begin{split}
\gamma_{\Omega_\pm}^{} &\;=\; \frac{1}{2}\Big( \Theta_1^{(\Omega_{\pm})} + \Theta_2^{\Omega_{\pm})} \Big)\, , \\[1mm]
 \Theta_1^{(\Omega_{\pm})} 
 &\;=\; \frac{\pi}{2}\,\frac{\Omega^2 - \Delta_1^2}{\Omega^2-\delta^2} \,S_1(\delta) \;+\; 
\frac{\pi}{2}\,\frac{\Delta_1^2-\delta^2}{\Omega^2-\delta^2}\,S_1(\Omega) \, , \\
 \Theta_2^{(\Omega_{\pm})} 
 &\;=\; \frac{\pi}{2}\,\frac{\Omega^2 - \Delta_2^2}{\Omega^2-\delta^2} S_2(\delta) \;+\; 
\frac{\pi}{2}\,\frac{\Delta_2^2-\delta^2}{\Omega^2-\delta^2}\,S_2(\Omega)  \,.
\end{split}
\]

As regards the relaxation rates $\gamma_5^{}$ and $\gamma_6^{}$ of $\langle\sigma_z\tau_z\rangle_t$, the situation is more subtle,
because they are determined by a quadratic equation in $\lambda$ which involves the self-energy $\Sigma_{1,2}(\lambda)$ 
in linear and second order in $K_{1,2}$. The calculation is most easily performed within the Redfield approach. The resulting
rate expressions are
\begin{equation}\label{ratesztz2}
 \begin{split}
\gamma_{5}^{} &\;=\; \frac{\Omega^2 -\Delta_1^2}{\Omega^2-\delta^2}\,\Theta_1^{(0)} 
                  +  \frac{\Omega^2 -\Delta_2^2}{\Omega^2-\delta^2}\,\Theta_2^{(0)} \, ,  \\[1mm]
\gamma_{6}^{} &\;=\; \frac{\Delta_1^2 -\delta^2}{\Omega^2-\delta^2}\,\Theta_1^{(0)} 
                  \;+\;  \frac{\Delta_2^2  - \delta^2}{\Omega^2-\delta^2}\,\Theta_2^{(0)} \, ,
\end{split}
\end{equation}
and the amplitudes read
\begin{equation}
B_{\gamma_5^{}} = \frac{v_z^2\delta^2}{(\Omega^2-\delta^2)^2}\;, \quad
B_{\gamma_6^{}} = \frac{v_z^2\Omega^2}{(\Omega^2-\delta^2)^2}\; .
\end{equation}
The functions $\Theta_{1,2}^{(0)}$ depend on the power spectra at the transition frequencies $\Omega$ 
and $\delta$. With the abbreviaton $\Delta_1^2 \pm \Delta_2^2 = \Delta_{\pm}^2$, we find the explicit form
\[
\begin{split}
\Theta_{1,2}^{(0)} &= \frac{\pi}{4}  \Big\{ S_{1,2}(\Omega) + S_{1,2}(\delta) \\
&\qquad \pm \, \frac{\Delta_-^4 + v_z^2\Delta_+^2}{(\Omega^2-\delta^2)\Delta_-^2}\,[S_{1,2}(\Omega)-S_{1,2}(\delta)]  \\
& \qquad \pm \, \frac{2v_z^2\Delta_{2,1}^2}{(\Omega^2-\delta^2)\Delta_-^2}\, [ S_{2,1}(\Omega)- S_{2,1}(\delta) ] \Big\}
\end{split}
\]
These expressions hold under assumption $\Delta_1\neq\Delta_2$.

\section{Summary}

We have studied the dynamics of a spin or qubit coupled to another spin. The latter could be another qubit, 
a bistable impurity, or a measuring device. We have given the dynamical equations in the WNR for general 
spin-spin coupling and we have discussed the rich features of the coupled dynamics. Analytic expressions for 
dephasing and relaxation rates and for the decay of the purity have been given in the one-phonon WNR limit.
Furthermore, the corresponding generalization to the quantum noise regime, which is based on a systematic 
calculation of the self-energy, has been presented. Our results smoothly match with those of the perturbative
Redfield approach. 

Financial support by the DFG through SFB/TR 21 is gratefully acknowledged.

\end{document}